\documentclass[runningheads]{llncs}
\usepackage[T1]{fontenc}
\usepackage{graphicx} 
\usepackage[ruled,linesnumbered]{algorithm2e}
\usepackage{graphicx}

\begin{document}
\title{A New Paradigm in Blockchain-based Financial Aid Distribution}

\author{Md. Raisul Hasan Shahrukh\inst{1} \and Md. Tabassinur Rahman\inst{2} \and Nafees Mansoor \inst{3}}

\institute{Department of Computer Science \& Engineering, \\ University of Liberal Arts Bangladesh, Dhaka-1207, Bangladesh \\
\email{raisul.hasan.cse@ulab.edu.bd}\inst{1}, \email{tabassenur.rahman.cse@ulab.edu.bd}\inst{2}, \email{nafees@ieee.org}\inst{3}
}

%

%
%
%
\maketitle              
\begin{abstract}
Blockchain technology has emerged as a game-changer in a variety of industries, providing robust solutions that can supplant conventional procedures. The unique potential of this technology originates from its decentralized ledger systems, which enable enhanced security, transparency, and the validation of transactions without the need for intermediaries. Notably, the financial sector is making substantial progress toward implementing blockchain solutions for a variety of operations, including remittances, lending, and investments. The healthcare industry is simultaneously incorporating this technology into systems for managing medical records, tracing supply chains, and data management. Similarly, the capacity of blockchain to enhance transparency, traceability, and accountability is widely acknowledged in supply chain management, from the procurement of basic materials to the delivery of finished goods. Diverse industries, including real estate, energy, and government, are actively investigating the potential of blockchain to improve efficiency, security, and transparency. Notably, Hyperledger Besu, an open-source blockchain platform, is used to implement smart contracts that automate processes and reduce manual intervention along distribution pathways. This exhaustive review examines the transformative potential of blockchain technology across a variety of industries, discussing the obstacles encountered and providing key insights into future research and development directions. This paper seeks to serve as a pivotal resource for academics, industry stakeholders, and policymakers by synthesizing existing scholarly literature and shedding light on significant findings.
\keywords{Blockchain \and Smart Contracts \and FinTech \and Process Automation \and Decentralized Ledger }
\end{abstract}

\section{Introduction}
Blockchain, a revolutionary technology conceived by Satoshi Nakamoto, underpins Bitcoin and has become a driving force in the financial technology (Fintech) sector \cite{b1}. The data structure, peer-to-peer network, and consensus mechanisms \cite{b2,b3} of this decentralized platform provide enhanced security, process simplification, and trust. Public, private, and consortium blockchains each have distinct advantages and disadvantages \cite{b4}. The burgeoning trend of blockchain implementation in the financial sector is revolutionizing fundamental processes. It improves cross-border transactions, trading, contract automation, auditability, and fraud detection, and has substantial implications for data security \cite{b5}. Consortium blockchains in particular establish a balance between privacy and transparency, making them ideal for financial applications \cite{b5}. This proposed system, a consortium blockchain-based decentralized application (DApp), exemplifies this development. It leverages smart contracts and robust digital identity verification\cite{b6} to facilitate efficient and transparent pension fund transfers in an effort to transform pension systems. This decentralized application reduces corruption risks and administrative costs while enhancing real-time auditability and accountability.

This document goes into detail on the significant influence blockchain technology has had on the banking industry. The distribution of funds in the humanitarian sector is covered in Section 2, while Section 3 goes into great length on the guiding principles of the suggested system. The system's design and execution are covered in Sections 4 and 5, respectively, while the blockchain revolution and its potential outcomes are covered in Section 4.

\section{Related Works}
Blockchain technology, specifically the consortium blockchain variant, is driving significant change across multiple industries, with the financial sector at the vanguard due to increased efficiency, cost savings, and enhanced security \cite{b7}. The technology redefines financial transactions by eliminating the need for intermediaries and fostering an auditable system as a result of its inherent immutability and transparency \cite{b7}. Financial institutions recognize the value of consortium blockchains, which strike a balance between privacy and transparency and are administered by a predetermined group of nodes \cite{b5}. This innovative approach is applicable not only to the financial sector, but also to a broader range of use cases, such as the development of intelligent cross-border transaction systems \cite{b5} and enhanced data privacy and security in health information systems \cite{b8}. Despite obstacles such as regulatory constraints, interoperability concerns, and user adoption issues, the potential benefits of consortium blockchain technology warrant additional investigation \cite{b9}. 

Within the context of stock exchange platforms, the transformative potential of consortium blockchain is especially notable. Al-Shaibani et al. Scalability and interoperability with existing stock exchanges are potential stumbling blocks for the technology. Consequently, future research should investigate innovative industry applications in these domains \cite{b1}. For secure data sharing in the healthcare industry, consortium blockchain systems such as HonestChain are proving effective \cite{b8}. The system combines the benefits of both public and private blockchains, enhancing security, privacy, and regulatory compliance. Similarly, consortium blockchain can considerably improve the sharing and storage of data in vehicular ad hoc networks (VANETs) \cite{b10}. Nevertheless, the ephemeral character of VANETs \cite{b11} and their limited computational resources \cite{b12}, present obstacles to the implementation of consortium blockchain. The proposed solutions include lightweight consensus algorithms \cite{b13}, mechanisms that protect privacy, and efficient data storage techniques \cite{b10}. In addition, consortium blockchain is making its imprint on intelligent transportation systems by providing secure data interchange and bespoke services \cite{b3}. It promises superior performance over traditional centralized systems in intelligent transportation \cite{b1}, mobile malware detection, and smart city governance \cite{b7}, resulting in enhanced decision-making and citizen engagement. The implications of consortium blockchain extend to smart home systems, where homomorphic encryption can provide privacy protection \cite{b14}. In the healthcare industry, which is beset by data security and access concerns, consortium blockchain provides a system that satisfies the diverse data access requirements of various stakeholders, thereby enhancing the efficiency and security of medical data management \cite{b15}.

Moreover, consortium blockchain technology bears promise for high-risk industries, such as coal extraction \cite{b16}, where it can facilitate data sharing, collaboration, and accountability among stakeholders, thereby enhancing safety. Consortium blockchain can improve data accountability and provenance monitoring \cite{b17}, which are crucial in sectors such as healthcare, finance, and governance. In the context of smart grids, consortium blockchain can resolve problems associated with a lack of trust and transparency \cite{b17}. Moreover, consortium blockchain technology proposes optimistic solutions for pension systems, facilitating efficiency, transparency, and secure pension access and administration \cite{b18}. Other possible applications include bolstering privacy protection in Vehicle-to-Grid (V2G) networks \cite{b19}, securing smart contracts, and protecting peer-to-peer file storage and sharing \cite{b20}. In conclusion, the vast potential of consortium blockchain technology is evident across numerous industries. It offers solutions to problems that conventional centralized systems struggle to address, indicating optimistic future developments and an imminent digital transformation.

\section{Proposed System }
This section describes the proposed platform, which aims to improve the transparency and efficacy of humanitarian aid distribution. The design, which is represented by a block diagram and Unified Modeling Language (UML) models, illustrates the system's essential components, their interactions, and object-oriented characteristics. These provide a clear comprehension of the structure and functionality of the system, laying the groundwork for future development and implementation.

\subsection{System Overview}
As depicted in Figure 1, the proposed system operates on a consortium blockchain involving diverse stakeholders including a fund-distribution entity, pension beneficiaries, and prospective financial service providers. Managed by the organizing entity, the primary node governs smart contracts, beneficiary enrollment, and fund distribution. Beneficiary nodes with the proper authorization can interact with the smart contract to verify balances or obtain funds.
\begin{figure}[htp]
    \centering
    \includegraphics[width=\linewidth]{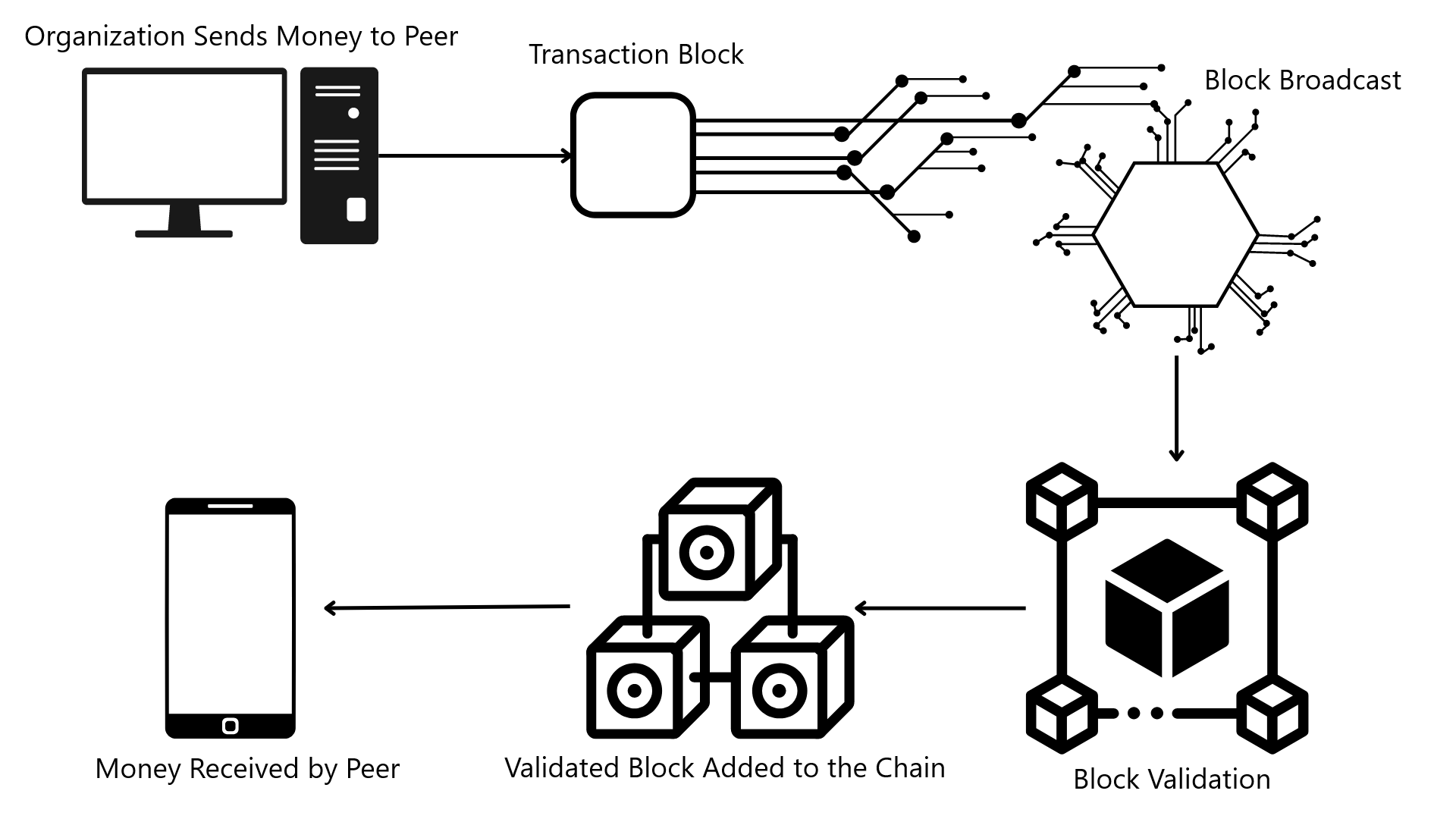}
    \caption{Proposed System's  Block Diagram}
    \label{fig:Proposed System Block Diagram}
\end{figure}
All organization-initiated transactions are securely recorded on the distributed ledger, ensuring security, auditability, and traceability. Integration with financial service providers for off-chain transactions improves the system's efficacy and control. This blockchain-based method guarantees a secure, transparent, and streamlined procedure for the distribution of funds.

\subsection{Technical Schematics}
The orderly workflow within this system is depicted in Fig. 2's activity diagram. It begins with the 'Organization' adding funds and managing recipients, a crucial stage for the fairness and integrity of the system. Afterwards, recipient bank accounts are registered, a crucial stage for ensuring accurate fund transfers and highlighting the system's emphasis on security. Provision of allowances to recipients concludes the procedure, contingent on the successful completion of preceding stages. 
\begin{figure}[htp]
    \centering
    \includegraphics[width=\linewidth]{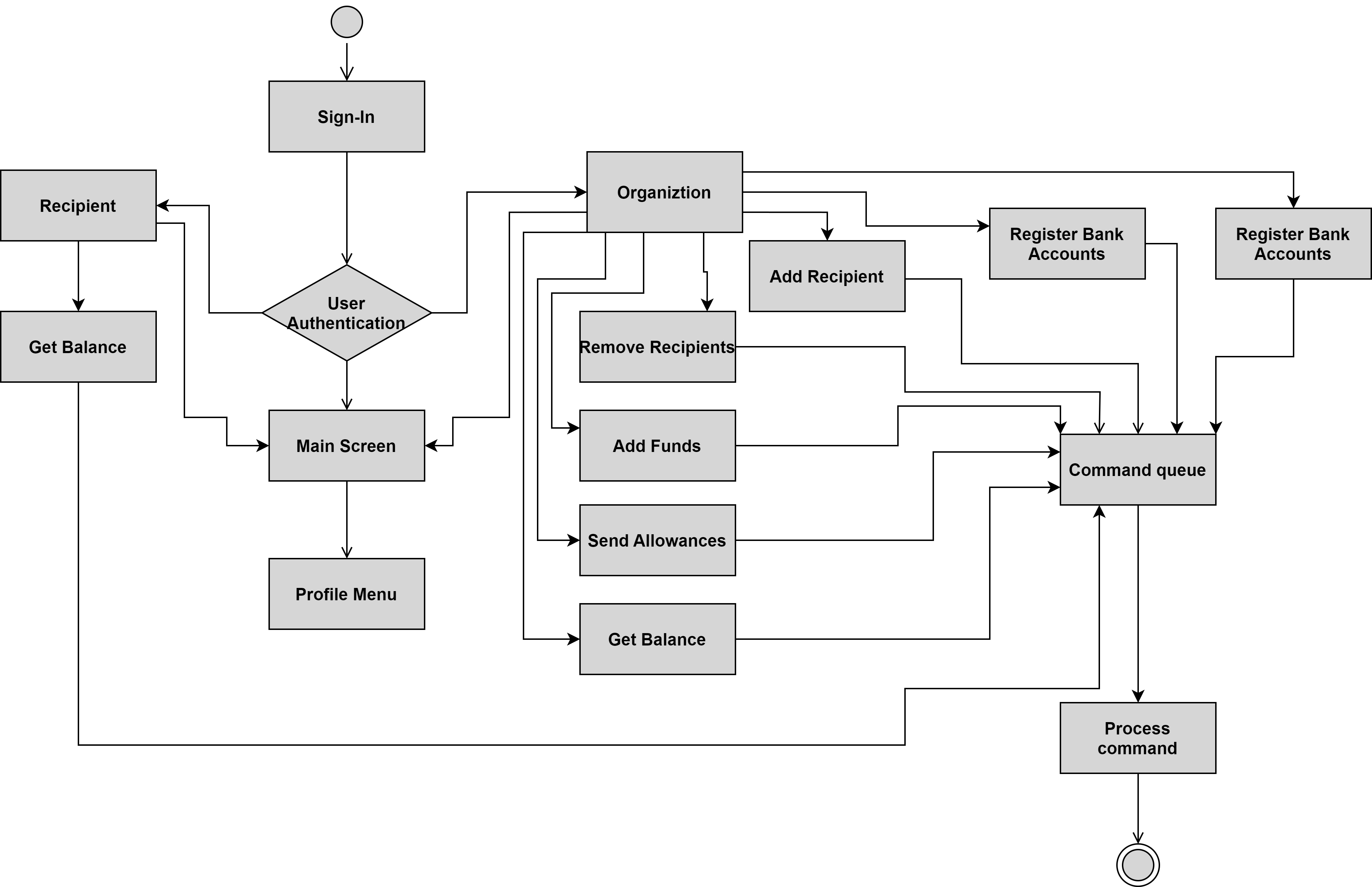}
    \caption{Proposed System's Activity Diagram }
    \label{fig:Proposed System Activity Diagram }
\end{figure}
\begin{figure}[htp]
    \centering
    \includegraphics[width=\linewidth]{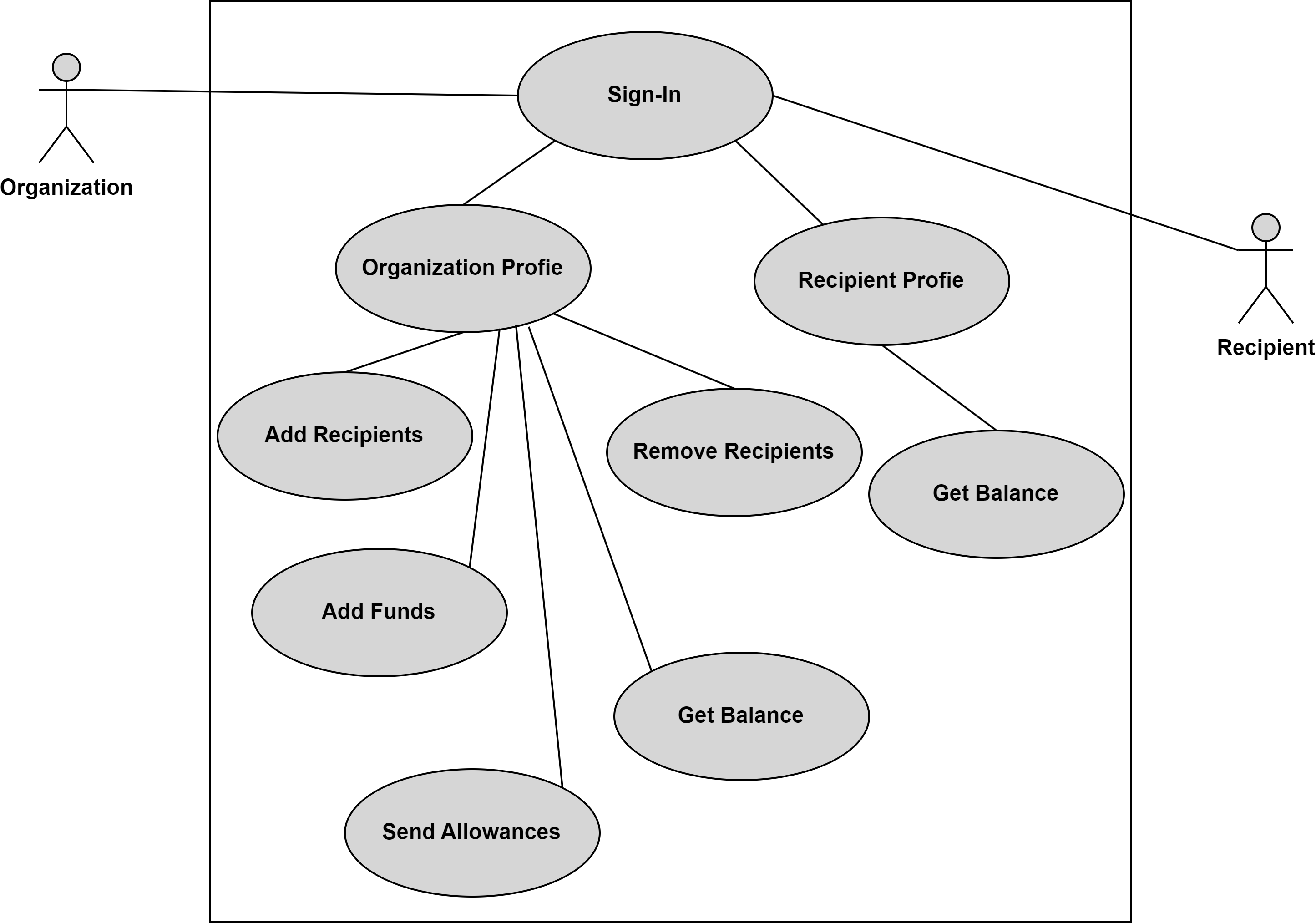}
    \caption{Proposed System's Use Case Diagram}
    \label{fig:Proposed System Use Case Diagram}
\end{figure}
\newpage
The sequential sequence, in which each action leads to the next, enhances the operational efficiency and dependability of the system. Consequently, the activity diagram encapsulates the interdependent system operations, indicating the system's overall dependability.

Use case diagram for the proposed system, illustrating how key actors interact with the system's features. The fact that the 'Organization' manages recipients, adds funds, distributes allowances, and registers bank accounts is indicative of its control over the system's operations. The 'Recipient,' although indirectly interacting with the system, is a crucial actor receiving allowances from the organization. The diagram effectively depicts the responsibilities and interactions of each actor, highlighting the well-structured design of the system. This diagram concisely summarizes the operation of this system, making complex processes simple to comprehend. Figure 3's use case diagram illustrates, in summary, a robust and structured method for aid distribution, creating the groundwork for an effective, transparent, and secure resource allocation process.

\section{Development \& Implementation of the Proposed System}
The suggested system uses the 'aNp' smart contract for transaction control and node interaction, which is supported by consortium blockchain technology. The 'organization' that has the contract manages the distribution of funds, and 'recipients' are the people who can use it. The distribution of pension funds is made secure, open, and effective thanks to the system's integration of smart contracts and off-chain transactions.

\subsection{Smart Contract}
The suggested method makes use of the Solidity-built aNp contract as a reliable mechanism for efficient and safe financial distribution within a consortium blockchain context. The authority to administer and distribute financial allowances is granted to an authoritative organization by this contract, which is deployed using the constructor function [Algorithm 1]. The AddRecipient and RemoveRecipient [Algorithm 1] methods are crucial for managing beneficiary lists for the organization. The RegisterBankAccount [Algorithm 1] function strengthens system security by ensuring secure connection of recipient addresses to bank accounts using keccak256 encryption.\newline
The company can distribute allowances thanks to the basic functionality of the aNp contract, which uses stringent guidelines for recipient authorization and fund availability checks to guarantee transactional security. The AddFunds function [Algorithm 2] in the contract also enables balance replenishment, enabling continued operation. Financial integrity is strengthened by the SendAllowance function [Algorithm 2], which guarantees secure fund transfers to authorized receivers. Finally, by enabling any address to check its balance,

\begin{algorithm}[H]
\LinesNotNumbered
\SetAlgoNoLine
\KwData{msg.sender as the current function caller.}
\KwResult{Constructor(), Initialize the contract with msg.sender as the organization, organization = msg.sender\;}
\LinesNotNumbered
\SetAlgoLined
\KwData{msg.sender as the current function caller, \_recipient as the address of the new recipient.}
\KwResult{Add a new recipient.}
 \If{msg.sender == organization}{
    recipients[\_recipient] = true\;
 }
\LinesNotNumbered
\SetAlgoLined
\KwData{msg.sender as the current function caller, \_recipient as the address of the recipient to be removed.}
\KwResult{Remove a recipient.}
 \If{msg.sender == organization}{
    recipients[\_recipient] = false\;
 }
\footnotesize
\LinesNotNumbered
\SetAlgoLined
\KwData{msg.sender: Current function caller}
\KwData{\_recipient: Recipient address, \_account: Account to be registered}
\KwResult{A registered bank account for a recipient.}
 \If{msg.sender == organization AND recipients[\_recipient] == true AND bytes(\_account).length > 0}{
    bankAccounts[\_recipient] = keccak256(abi.encodePacked(\_account))\;
    emit BankAccountRegistered (\_recipient, bankAccounts[\_recipient])\;
 }
\caption{Authorization Functions}
\end{algorithm}

\begin{algorithm}[H]
\LinesNotNumbered
\SetAlgoLined
\KwData{msg.sender as the current function caller, \_amt as the amount to be added.}
\KwResult{Add funds to the organization's balance.}
 \If{msg.sender == organization}{
    balances[organization] += \_amt\;  emit FundsAdded(msg.value)\;
 }
\LinesNotNumbered
\SetAlgoLined
\KwData{msg.sender as the current function caller, \_recipient as the address of the recipient, \_amount as the amount to be sent.}
\KwResult{Send allowance to a recipient.}
 \If{msg.sender == organization AND recipients[\_recipient] == true AND balances[organization] >= \_amount}{
    balances[organization] -= \_amount\;
    emit AllowanceSent(\_recipient, \_amount)\;
 }
\SetAlgoLined
\KwData{msg.sender as the current function caller.}
\KwResult{Return the balance of the function caller, return balances[msg.sender]\;}
\caption{Transaction Functions}
\end{algorithm}
the GetBalance  function [Algorithm 2] enables transparency. These characteristics come together to create a safe, traceable, and transparent mechanism for financial distribution.

\subsection{Smart Contract Functionality}
Written in Ethereum's Solidity programming language, the "aNp" smart contract functions as the system's backbone. The contract establishes the address of the organization and maintains listings of recipients, their balances, and encrypted bank information \cite{b14}. The contract initialization function identifies the organization as the deploying address, thus identifying the system administrator \cite{b18}. Important operations, including recipient administration and fund distribution, are restricted to this address by the 'onlyOrganization' modifier. 'addRecipient','removeRecipient', and'sendAllowance' functions facilitate recipient list modifications and disbursement of funds, assuming adequate balance \cite{b15}. 

The 'addFunds' function allows the organization to add funds to a contract, whereas the 'getBalance' function allows any participant to view their balance \cite{b21}. The'registerBankAccount' function securely stores recipients' hashed bank account information \cite{b3}. Important operations generate event logs for auditability and transparency.

\subsection{Private Decentralized Network}
Hyperledger Besu is an enterprise-grade, open-source blockchain platform created by the Hyperledger consortium. It supports both public and private Ethereum-compatible networks and provides a flexible foundation for the development and deployment of decentralized applications. Using Hyperledger Besu, we have effectively deployed a private consortium blockchain network in this instance. 

This private network ensures that only authorized users have access to a controlled and secure environment. Using Truffle, a comprehensive development environment, testing framework, and asset infrastructure for Ethereum, we facilitated the efficient deployment of our smart contract during the setup process.

Significant is the successful deployment [See Fig.4] of our system on a private Hyperledger Besu network. It provides numerous advantages, including transaction secrecy, data privacy, and the ability to leverage existing Ethereum infrastructure, all of which are essential for business applications. It improves the system's overall robustness, scalability, and security, assuring a reliable platform for secure transactions and interactions, thereby improving the system's overall functionality.
\begin{figure}[htp]
    \centering
    \includegraphics[width=.86\linewidth]{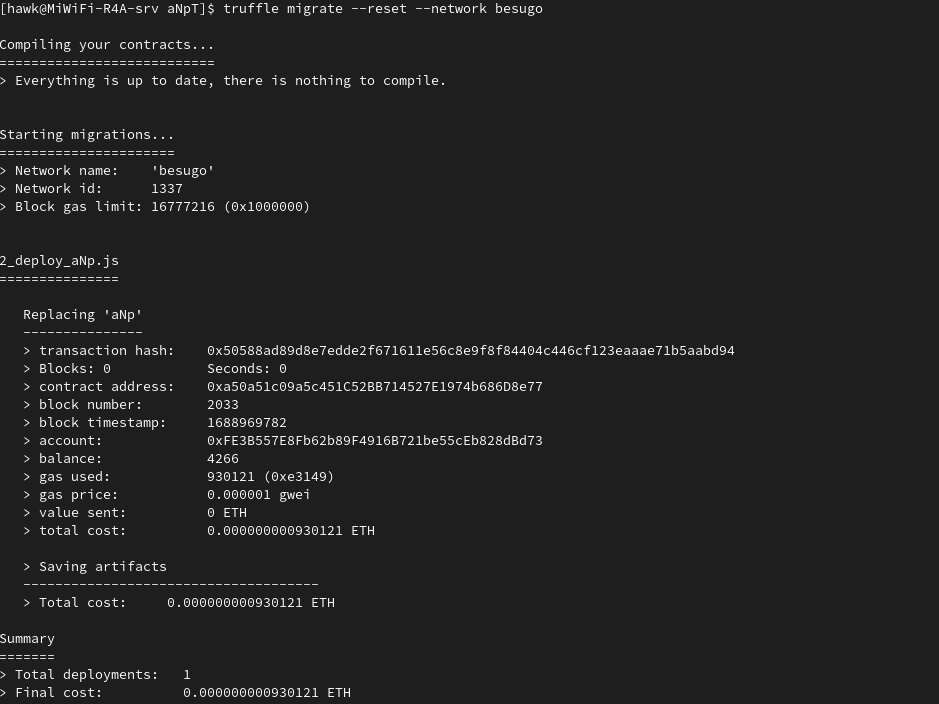}
    \caption{Smart Contract Deployment on HyperLedger Besu Private Network }
    \label{fig:Smart Contract Deployment on HyperLedger Besu Private Network  }
\end{figure}
\newpage
\section{Conclusion and Future Works} 
The unique combination of transparency and privacy in consortium blockchain technology presents significant opportunities for sector-wide transformation. Despite difficulties with scalability and interoperability, it is imperative to investigate its potential and establish robust regulatory structures. Future research should focus on enhancing interoperability, scalability, and the development of smart contracts with the potential incorporation of AI. Additionally, we must comprehend the socioeconomic ramifications of widespread adoption and the government's function in drafting an accommodating legal framework. Such a comprehensive and multifaceted research strategy would allow us to fully exploit the potential of consortium blockchain technology.


\begin{thebibliography}{00}

\bibitem{b1} Al-Shaibani, H., Lasla, N., \& Abdallah, M. (2020). “Consortium blockchain-based
decentralized stock exchange platform” IEEE Access, 8, 123711–123725

\bibitem{b2} Al Sunaidi, S. J., \& Alhaidari, F. A. (2019). A survey of consensus algorithms for blockchain technology. In 2019 international conference on computer and information
sciences (iccis) (pp. 1–6).

\bibitem{b3} Wang, D., \& Zhang, X. (2020). Secure data sharing and customized services for
intelligent transportation based on a consortium blockchain. IEEE Access, 8,
56045–56059.

\bibitem{b4} Fan, M., \& Zhang, X. (2019). Consortium blockchain based data aggregation and
regulation mechanism for smart grid. IEEE Access, 7, 35929–35940

\bibitem{b5} Fu, Z., Dong, P., Li, S., \& Ju, Y. (2021). An intelligent cross-border transaction system based on consortium blockchain: A case study in shenzhen, china. Plos one, 16(6), E0252489.

\bibitem{b6} Rahman, T., \& Mouno, S. I., \& Raatul, A. M., \& Azad, A. K. A., \& Mansoor, N. (2023). Verifi-Chain: A Credentials Verifier using Blockchain and IPFS. arXiv preprint arXiv: 2307.05797.

\bibitem{b7} Bai, Y., Hu, Q., Seo, S.-H., Kang, K., \& Lee, J. J. (2021). Public participation
consortium blockchain for smart city governance. IEEE Internet of Things Journal,
9(3), 2094–2108.

\bibitem{b8} Purohit, S., Calyam, P., Alarcon, M. L., Bhamidipati, N. R., Mosa, A., \& Salah, K.(2021). Honestchain: Consortium blockchain for protected data sharing in health
information systems. Peer-to-peer Networking and Applications, 14(5), 3012–3028.

\bibitem{b9} Srivastav, R. K., Agrawal, D., Shrivastava, A., et al. (2020). A survey on vulnerabilities and performance evaluation criteria in blockchain technology.

\bibitem{b10} Zhang, X., \& Chen, X. (2019), Data security sharing and storage based on a
consortium blockchain in a vehicular ad-hoc network. IEEE Access, 7, 58241–
58254.

\bibitem{b11} Akter, S., \& Mansoor, N. (2020). A spectrum aware mobility pattern based routing protocol for CR-VANETs. 2020 IEEE Wireless Communications and Networking Conference (WCNC), 1--6.

\bibitem{b12} Mansoor, N., \& Hossain, M. I., \& Rozario, A., \& Zareei, M., \& Arreola, A. R. (2023). A Fresh Look at Routing Protocols in Unmanned Aerial Vehicular Networks: A Survey.IEEE Access.

\bibitem{b13} Onee, B. A. H., \& Antora, K. F., \& Rajme, O. S., \& Mansoor, N. (2023). Development of a Blockchain-Based On-Demand Lightweight Commodity Delivery System.  arXiv, cs.DC, 2307.08050.

\bibitem{b14} She, W., Gu, Z.-H., Lyu, X.-K., Liu, Q., Tian, Z., \& Liu, W. (2019). Homomorphic consortium blockchain for smart home system sensitive data privacy preserving.
IEEE Access, 7, 62058–62070.

\bibitem{b15} Xu, B., Xu, L. D., Wang, Y., \& Cai, H. (2022). A distributed dynamic authorisation method for internet+ medical \& healthcare data access based on consortium blockchain. Enterprise Information Systems, 16(12), 1922757.

\bibitem{b16} Qiang, Z., Wang, Y., Song, K., \& Zhao, Z. (2021). Mine consortium blockchain: the application research of coal mine safety production based on blockchain. Security
and Communication Networks, 2021.

\bibitem{b17} Neisse, R., Steri, G., \& Nai-Fovino, I. (2017). A blockchain-based approach for data accountability and provenance tracking. In Proceedings of the 12th international
conference on availability, reliability and security (pp. 1–10).

\bibitem{b18} ANDRIANOVA, A., HAUNER, P., MCDONALD, D. K., MANNING, D. A., \&
ZEROUALI, M. (n.d.). Akropolis: A global blockchain pensions infrastructure.

\bibitem{b19} Wei, G., \& Ma, Y. (2021). Privacy protection strategy of vehicle-to-grid network based on consortium blockchain and attribute-based signature. In Iop conference series: Earth and environmental science (Vol. 661, p. 012027).

\bibitem{b20} Peng, S., Bao, W., Liu, H., Xiao, X., Shang, J., Han, L., . . . Xu, Y. (2023). A peer-to-peer file storage and sharing system based on consortium blockchain. Future Generation Computer Systems, 141, 197–204.

\bibitem{b21} Yue, K., Zhang, Y., Chen, Y., Li, Y., Zhao, L., Rong, C., \& Chen, L. (2021). A survey of decentralizing applications via blockchain: The 5g and beyond perspective. IEEE Communications Surveys \& Tutorials, 23(4), 2191–2217.


\end{thebibliography}
\end{document}